\documentstyle[prl,epsf,aps,twocolumn]{revtex}
\def\gsim{\mathrel{\rlap {\raise.5ex\hbox{$ > $}}
{\lower.5ex\hbox{$\sim$}}}}
\def\lsim{\mathrel{\rlap {\raise.5ex\hbox{$ < $}}
{\lower.5ex\hbox{$\sim$}}}}

\newcommand{\be}{\begin{equation}}
\newcommand{\ee}{\end{equation}}
\newcommand{\bea}{\begin{eqnarray}}
\newcommand{\nn}{\nonumber}
\newcommand{\eea}{\end{eqnarray}}

\newcommand{\C}{C_\epsilon}
\newcommand{\D}{D_\epsilon}
\newcommand{\ggg}{2\eta\epsilon^2 \log|z/a|^2}
\def\gappeq{\mathrel{\rlap {\raise.5ex\hbox{$>$}}
{\lower.5ex\hbox{$\sim$}}}}
 
\def\lappeq{\mathrel{\rlap{\raise.5ex\hbox{$<$}}
{\lower.5ex\hbox{$\sim$}}}}
 
\begin{document}
\title { \bf  A Microscopic Recoil Model for Light-Cone Fluctuations
in Quantum Gravity}
\author{John Ellis}
\address{Theory Division, CERN, CH-1211 Geneva 23, Switzerland}
\author{N.E. Mavromatos}
\address{Department of Physics, University of Oxford, 
1 Keble Road,  
Oxford OX1 3NP, U.K., and \\Theory Division, CERN, 
CH-1211 Geneva 23, Switzerland}

\author {D.V. Nanopoulos}
\address{Department of Physics, 
Texas A \& M University, College Station, 
TX~77843-4242, USA,
Astroparticle Physics Group, Houston
Advanced Research Center (HARC), Mitchell Campus,
Woodlands, TX 77381, USA, and 
Academy of Athens, \\
Chair of Theoretical Physics, 
Division of Natural Sciences, 28~Panepistimiou Avenue, 
Athens 10679, Greece.}
\address{\mbox{ }}
\address{\parbox{14cm}{\rm \mbox{ }\mbox{ }
We present a microscopic model for light-cone
fluctuations {\it in vacuo}, which incorporates a treatment of
quantum-gravitational recoil effects induced by energetic particles.
Treating defects in space-time as solitons in string theory,
we derive an energy-dependent
refractive index and a stochastic spread in
the arrival times of mono-energetic photons due to
quantum diffusion through space-time foam,
as found previously using an
effective Born-Infeld action. Distant astrophysical
sources provide sensitive tests of these possible
quantum-gravitational phenomena.
}}
\address{\mbox{ }}
\address{\parbox{14cm}{\rm PACS No:04.60.-m, 04.62.+v~~~~ACT-4/99~~~~
CTP-TAMU-24/99~~~~OUTP-99-25P~~~~gr-qc/9906029}}
\maketitle

The theory of quantum gravity is still elusive,
despite the progress made in recent years in 
various different approaches, including those based on canonical
field theory~\cite{ashtekar} and on the theory formerly known as
string~\cite{dbranes}.
Quantum fluctuations in the
underlying space-time, such as might lead to
the formation and evaporation 
of microscopic black holes, are still not understood,
although the modern representation of
defects in space time as string solitons has
yielded deep insights into massive black holes in
flat space-time.

Quantum-gravitational fluctuations in the space-time background might
cause the vacuum to have non-trivial optical properties,
due in particular to recoil induced by a
propagating energetic particle. 
An important issue in 
the study of such effects is whether
manifest Lorentz invariance is retained.
In our view~\cite{nature,emnnew}, 
as well as that of others~\cite{ashtekar,pullin,ford,garay},
a quantum-gravity ground state may not be
Lorentz invariant
since the quantization of the 
metric field about a {\it given} background
may induce a preferred frame. This could be regarded as
a spontaneous breaking of Lorentz invariance, which is
probably a necessary step in viewing the
gravitational vacuum as a `medium'. Such a symmetry
breakdown is inevitable in the canonical approach to
quantizing the gravitational field, which
leads to quantum fluctuations in the light-cone.
We and others have suggested that the breakdown
of Lorentz invariance may lead to
observable consequences for the propagation of
light~\cite{nature,emnnew,pullin}, and the 
main purpose of this letter is to
derive these via a microscopic recoil model for
light-cone fluctuations.

Light propagating through media with
non-trivial optical properties may
exhibit a frequency-dependent refractive
index, namely a variation in the light
velocity with photon energy. Another possibility is
a difference between the velocities of light with
different polarizations, namely birefringence,
and a third is a diffusive spread in the apparent velocity of
light for light of fixed energy (frequency). We have derived 
the first~\cite{nature} and third~\cite{emnnew} effects previously using a
formal
approach based on a Born-Infeld Lagrangian obtained
from solitonic effects in string theory, and the
possibility of birefringence has been raised~\cite{pullin} within a
canonical approach to quantum gravity. A different approach
to light propagation has been taken in~\cite{ford}, where
quantum-gravitational fluctuations in the light-cone have been
calculated. In this paper, we use this formalism together with
a microscopic model background obtained from considering the
quantum recoil of a string soliton
to derive a non-trivial refractive index and a diffusive spread
in the arrival times of photons of given frequency. 

We first review briefly the analysis in \cite{ford},
which considered gravitons in a squeezed coherent state, the natural result 
of quantum creation in the presence of black holes.  
Such gravitons induce quantum fluctuations in the space-time metric, 
in particular fluctuations in the light-cone~\cite{ford},
i.e., stochastic fluctuations in the 
velocity of light propagating through this `medium',
which may have observable effects $\Delta t$ on the arrival times of
photons. Following~\cite{ford},
we consider a flat background space-time 
with a linearized perturbation, corresponding to the 
invariant metric element
$ds^2=g_{\mu\nu}dx^\mu dx^\nu = 
\left(\eta_{\mu\nu} + h_{\mu\nu}\right)dx^\mu dx^\nu 
= dt^2 - d{\overline x}^2 + h_{\mu\nu}dx^\mu dx^\nu $.
Let 
$2 \sigma (x,x')$ be the squared geodesic separation 
for any pair 
of space-time 
points $x$ and $x'$, and let $2 \sigma_0(x,x')$ denote the corresponding 
quantity in a flat space-time background. 
In the case of small 
gravitational perturbations about the flat background,
one may expand
$\sigma = \sigma _0 + \sigma_1 + \sigma_2 + \dots$,
where $\sigma_n$ denotes the $n$-th order term in 
an expansion in the gravitational perturbation $h_{\mu\nu}$.
Then, as shown in \cite{ford}, 
the root-mean-square (RMS) deviation from the classical propagation 
time $\Delta t$ is related to $<\sigma^2>$ 
by:
\be
   \Delta t = \frac{\sqrt{<\sigma^2> -
<\sigma_0^2>}}{L} \simeq \frac{\sqrt{<\sigma_1^2>}}{L} + \dots 
\label{deltat}
\ee
where $L = |x' - x|$ is the distance between the source and the detector.
The expression (\ref{deltat}) is gauge invariant~\cite{ford}.

The model we propose in this paper is based on solitons in
string theory, which are described formally as $D$ branes~\cite{dbranes}. 
As
commented earlier, one may expect Lorentz invariance to be
broken in a generic theory of quantum gravity. In the
context of string theory, this entails the exploration of
non-critical string backgrounds, since Lorentz invariance is 
related to the conformal symmetry that is a
property of critical strings. A general approach to the formulation of
non-critical string theory involves introducing a Liouville field~\cite{ddk}
as a conformal factor
on the string world sheet, which has non-trivial dynamics and
compensates the non-conformal behaviour of the string background.
In the case of $D$-brane string solitons, their recoil after
interaction with a closed-string state~\cite{kmw}, as used to describe a
photon, is characterized by a pair of logarithmic operators~\cite{gur}:
\be
C_\epsilon \sim \epsilon \Theta_\epsilon (t),\qquad
D_\epsilon \sim t \Theta_\epsilon (t)
\label{logpair}
\ee
defined on the boundary $\partial \Sigma$ of the string
world sheet.

The operators (\ref{logpair}) act
as deformations of the conformal field theory on the
world sheet: $U_i \int _{\partial \Sigma} \partial_n X^i \D$ 
describes the shift of the $D$
brane induced by the scattering, where
$U_i$ is its recoil velocity, and $ Y_i 
\int _{\partial \Sigma} \partial_n X^i \C$
describes quantum fluctuations in the initial
position $Y_i$ of the $D$ particle. It has been
shown~\cite{ms} that
energy-momentum is conserved during the recoil process:
$U_i = k_1 - k_2$, where $k_1 (k_2)$ is the momentum of 
the propagating closed-string state before (after) the recoil,
as a result of the summation over world-sheet genera.
We also note that $U_i = g_s P_i$, where $P_i$ 
is the momentum and $g_s$ is the string coupling,
which is assumed here to be weak enough
to ensure that $D$ 
branes are very massive, with mass $M_D=1/(\ell _s g_s)$,
where $\ell _s$ is the string length.   

The correct specification of the logarithmic pair (\ref{logpair})
entails a regulating 
parameter $\epsilon \rightarrow 0^+$, which
appears inside the $\Theta (t)$ operator:
$\Theta (t) = \int \frac{d\omega}{2\pi}\frac{1}{\omega -i\epsilon} 
e^{i\omega t}  $. In order to realize
the logarithmic algebra between the operators $C$ and $D$,
one takes~\cite{kmw}:
$\epsilon^{-2} \sim {\rm Log}\Lambda/a \equiv \alpha$,
where $\Lambda$ ($a$) are infrared (ultraviolet) world-sheet cut-offs. 
The recoil operators (\ref{logpair}) are slightly relevant,
in the sense of the renormalization group for the
world-sheet field theory, with small
conformal dimensions $\Delta _\epsilon = -\frac{\epsilon^2}{2}$. 
The relevant two-point functions have the following form: 
\bea
&~&<\C(z)\C(0)> \stackrel{\epsilon\to 0}{\sim} 0+
{\cal O}(\epsilon^2) \nn \\
&~&<\C(z)\D(0)> 
\stackrel{\epsilon\to 0}{\sim} {\pi\over2}\sqrt{{\pi\over\epsilon^2\alpha}}
\left(1-\ggg\right) \nn \\
&~&<\D(z)\D(0)> =\frac{1}{\epsilon^2} <\C(z)\D(0)> \nn \\
&~&\stackrel{\epsilon\to 0}{\sim} {\pi\over2}\sqrt{{\pi\over\epsilon^2\alpha}}
\left({1\over\epsilon^2}-2\eta\log|z/a|^2\right)
\label{twopoint}
\eea
which is the logarithmic algebra~\cite{gur} 
in the limit $\epsilon \rightarrow 0^+$,
modulo the leading divergence in the $<\D\D>$ recoil correlator.
In fact, it is this leading divergent term 
that will be of importance for our purposes in 
this article.  

It is the fact that the recoil operators are relevant
operators in a 
world-sheet renormalization-group sense that
requires dressing the world-sheet theory with a Liouville
field~\cite{ddk}
in order to restore conformal invariance, which has been lost in the 
recoil process. One then makes the crucial step of 
identifying 
the world-sheet zero mode of the Liouville field with the target
time $t$, which is justified in~\cite{emndbrane,ms}
using the logarithmic algebra (\ref{twopoint}) for the case at hand. 
This identification leads to the appearance of a
curved space-time background in target space, with 
the metric elements~\cite{kanti} 
\be
G_{ij} =\delta _{ij} , G_{00}=-1, G_{0i}=\epsilon(\epsilon Y_i + U_i t)\Theta _\epsilon (t) 
\label{metrictarget}
\ee
where the suffix $0$ denotes temporal (Liouville) components. 

We now remark~\cite{ms} that the velocity operator 
$\D$ (\ref{logpair}) becomes exactly marginal,
in a world-sheet renormalization-group sense, upon
division by $\epsilon $, in which case the recoil
velocity is renormalized:
$U_i \rightarrow {\overline U}_i \equiv U_i /\epsilon $
is the physical recoil velocity. 
Viewed as a perturbation about a flat target space-time, 
the metric (\ref{metrictarget}) implies that 
that the only non-zero components of $h_{\mu\nu}$
are:
\be
h_{0i} = \epsilon ^2 {\overline U}_i t \Theta _\epsilon (t)
\label{pert2}
\ee
in the case of $D$-brane recoil.

Consider light propagation along the $x$ direction in
the presence of a metric fluctuations $h_{0x}$ (\ref{pert2})
in flat space, along a null geodesic
given by $(dt)^2 = (dx)^2 + 2 h_{0x} dt dx $.
For large times $t \sim {\rm Log}\Lambda/a
\sim \epsilon^{-2}$~\cite{kanti}, $h_{ox} \sim
{\overline U}$, and thus we obtain
\be
\frac{cdt}{dx}={\overline U} + \sqrt{1 + {\overline U}^2} \sim 
1 + {\overline U} + {\cal O}\left({\overline U}^2\right)
\label{refr}
\ee
where the recoil velocity $\overline U$ is in the direction of the
incoming
light ray. Taking into account energy-momentum 
conservation  in the recoil process, 
which has been derived in this formalism as mentioned previously, 
one has a typical order of magnitude 
${\overline U}/c ={\cal O}(E/M_Dc^2)$,
where $M_D =g_s^{-1}M_s$ is the $D$-brane mass scale, with $M_s \equiv 
\ell
_s^{-1}$. Hence (\ref{refr}) implies 
a subluminal energy-dependent velocity of light:
\be
c(E)/c=1 -{\cal O}\left(E/M_Dc^2\right)
\label{vellight}
\ee
which corresponds to a {\it classical} refractive index.
This appears because the metric perturbation (\ref{pert2}) 
is energy-dependent,
through its dependence on ${\overline U}$.

The subluminal velocity (\ref{vellight})
induces a delay in the arrival of a photon of energy $E$
propagating over a distance $L$
of order: 
\be
      (\Delta t)_r =  \frac{L}{c}{\cal O}\left(\frac{E}{M_Dc^2}\right)
\label{figmerclass}
\ee
This effect can be understood physically from the fact  
that the curvature of space-time induced by the recoil is
${\overline U}-$ and hence energy-dependent.
This affects the paths of photons
in such a way that more energetic photons see more
curvature, and thus are delayed with respect to low-energy ones.

The absence of superluminal light propagation
was found previously
via the formalism of the Born-Infeld 
lagrangian dynamics of $D$ branes~\cite{ms,emnnew}.
Furthermore, the result (\ref{figmerclass}) is in agreement 
with the analysis of~\cite{aemn,nature}, 
which was based on a more abstract analysis of Liouville strings.
It is encouraging that this follows also from a more conventional 
general relativity approach~\cite{ford}, in which the underlying
physics is transparent. We find no birefringence effect,
in contrast to~\cite{pullin}.

As a preliminary to evaluating
{\it quantum effects}~\cite{emnnew}
in the context of our simplified 
one-dimensional problem, we now express
$\sigma_1^2$ in terms
of the two-point function of $h_{0x}$,
considering again the null geodesic in the presence of the small metric 
perturbations (\ref{pert2}). To first order in the recoil velocity,
one has: 
$c(\Delta t) = \left(\Delta x  - 2\int _{x_1}^{x_2} dx 
h_{0x}\right)$,  
from which we find
$2\sigma \simeq  (\Delta t)^2 - 
(\Delta x)^2 + 2\epsilon^2 {\overline U}t 
\Theta _\epsilon (t) (\Delta x)^2 $, implying that
$\sigma _1 = \epsilon ^2 {\overline U} t \Theta _\epsilon (t)$.   
One then has:
\be
<\sigma _1^2 > \sim  L^2 \int _x^{x'} \int _x^{x'}
dy \int dy' <h_{0x}(y,t)h_{0x}(y',t')>
\label{metric}
\ee
In the case of $D$-brane recoil, the computation 
of the quantum average $< \dots >$ may be made in
the Liouville-string approach
described above. In this case, the quantum average 
$< \dots >$ is replaced by a world-sheet correlator calculated
with a world-sheet action deformed by (\ref{logpair}).  
It is clear from (\ref{pert2}) that 
the two-point metric correlator appearing in (\ref{metric})
is just the $<\D\D>$ world-sheet recoil 
two-point function described in (\ref{twopoint}). 
Thus, at the classical tree level on the world sheet,
one may recover the refractive index
(\ref{figmerclass}) from (\ref{metric}) by concentrating
on the leading divergence in the $\D \D$ correlator
(\ref{twopoint}), proportional to $\epsilon ^{-2}$.

To describe fully the quantum effects, one must
sum over world-sheet genera. As discussed in~\cite{emndbrane,ms} 
such a procedure results in a canonical quantization 
of the world-sheet couplings, which, in the problem at hand,
coincide with the target-space collective coordinates and momenta
of the recoiling $D$ brane. Thus, the quantum effects arising from 
summation over world-sheet genera result, in a first approximation, in
$<\sigma_1^2 > \sim L^4 \left( \Delta {\overline U}\right)^2 $,
where $\Delta {\overline U}$ denotes the quantum uncertainty 
of the recoil velocity ${\overline U}$, which has been computed 
in~\cite{ms}.  

The result of the summation over genera has been discussed
in detail in~\cite{emndbrane,ms}, and is not repeated here: we only state
the final results relevant for our purposes. 
The leading contributions 
to the quantum fluctuations in the
space ${\cal M}$ of world-sheet couplings
arise from pinched annulus diagrams 
in the summation over world-sheet general~\cite{emndbrane,ms}. 
These consist of thin tubes of width $\delta\rightarrow 0$, which
one may regard as wormholes attached to 
the world-sheet surface $\Sigma$. The
attachment of each tube corresponds to the insertion of a bilocal pair
of recoil vertex operators 
$V^i(s)V^j(s')$  on the boundary
$\partial\Sigma$, with interaction strength
$g_s^2$, and computing the string propagator along the thin tubes. 
One effect of the dilute gas of world-sheet wormholes
is to exponentiate the bilocal operator, leading to a
change in the world-sheet action~\cite{emndbrane,ms}. This
contribution can be cast into the form of a local action by rewriting it as a
Gaussian functional 
integral over wormhole parameters $\rho_i^{ab}$, leading
finally to~\cite{ms}:
\be
\sum_{\rm genera}Z \simeq 
\int_{\cal
M}D\rho~e^{-\rho_i^{ab}G^{ij}\rho_j^{cd}/2|\epsilon|^2\ell_s^2
g_s^2\log\delta}~\left\langle W[Y+\rho]\right\rangle_0
\label{genusexp}\ee
where $\langle W[Y+\rho]\rangle_0$
denotes the partition function on a world sheet with the topology of a
disc of a model deformed by the operators (\ref{logpair}),
with couplings shifted by $\rho$, and
$G^{ij}$ is the inverse of the 
Zamolodchikov metric~\cite{zam}
$<V_iV_j>$ in string theory space, evaluated 
on the disc world sheet.
The shifts in the effective couplings $Y_i,U_i$  
imply that they fluctuate statistically,
in much the same way as wormholes and other topology changes in target
space times
lead to the quantization of the couplings in conventional field 
theories~\cite{coleman}. 

In our case, we see from (\ref{genusexp}) that the effect of this
resummation over pinched
genera is to induce quantum fluctuations in the solitonic $D$-brane
background, giving rise to a
Gaussian statistical spread in the 
collective coordinates of the $D$ brane,  
determined by $G^{ij}$ 
for the logarithmic deformations (\ref{logpair}).
Note that, in such a formalism,
one defines a renormalized string coupling ${\bar g}_s = g_s \epsilon
^{-1}$,
which plays the r\^ole of the physical string coupling in the 
problem~\cite{ms}.   
To lowest non-trivial order in the string coupling ${\bar g}_s$
and ${\overline U}^2$, the 
world-sheet analysis
of~\cite{ms}, using the methods of logarithmic 
conformal field theory~\cite{gur}, showed that the quantum fluctuations 
in ${\overline U}$ induced by the summation
over world-sheet topologies are given by;
\be
\left( \Delta {\overline U} \right)^2  = 4 {\bar g}_s^2
\left[1 - \frac{285}{2}{\overline U}^2\right],
\label{delU}
\ee
if the $D$-brane foam corresponds to a 
minimum-uncertainty wave-packet~\cite{ms}, with the position fluctuations
of the $D$ branes
being saturated: $\Delta Y_i \sim {\bar g_s}^{1/3}\ell_s$.  
The energy-independent first part of (\ref{delU}),
can be absorbed into $\sigma_0$, and hence
does not contribute to the stochastic 
fluctuations (\ref{deltat}) in the photon arrival time.
The second part of the 
uncertainty (\ref{delU}), which depends on $\overline U$,
and hence the energy of the photon,
cannot be scaled away for all photons of different energies 
by a simple coordinate transformation. 
However, it is of higher 
(second) order in the small parameter ${\overline U}$.
Thus the geodesic correction $\sigma_1$, which is linear in the
gravitational
perturbation $h_{\mu\nu}$, leads in leading order only to a classical 
refractive index.

We now repeat the analysis for the quadratic correction $\sigma_2$
in (\ref{deltat}). This
contains the normal-ordered coincidence limit 
of the two-point function of the $D_\epsilon$ operator, which
has a leading-order term
proportional to $\epsilon ^{-2}$ times the $c$-number identity,
as seen in (\ref{twopoint}). Since we work in 
a subtraction scheme in which
one-point functions of the deformations 
(\ref{logpair}) vanish, this means that
$<\sigma_1 \sigma_2> =0$, so that the leading 
contribution comes from the sum over genera of $<\sigma_2^2>$, which
yields fluctuations in the arrival time that are
proportional to the uncertainty $\Delta \left({\overline U}^2\right)
\sim {\overline U}\Delta {\overline U}$.
For a minimum-uncertainty 
state of $D$ branes, we therefore find a contribution 
\be 
| \left( \Delta t\right)_{obs}| \simeq 
{\cal O}\left({\bar g}_s\frac{E}{M_D c^2}\right) \frac{L}{c} 
\label{figuremerit}
\ee
to the RMS fluctuation in arrival times.
As expected, the {\it quantum} effect (\ref{figuremerit}) 
is suppressed by a power of the string coupling constant, 
when compared with the 
classical refractive index effect (\ref{figmerclass}).
The result (\ref{figuremerit}) 
was derived in~\cite{emnnew} using 
the techniques of Liouville string theory,
via the Born-Infeld lagrangian for the 
propagation of photons in the $D$-brane foam. 
Its derivation here using more
conventional field-theoretic techniques adds robustness to the 
Liouville string theory perspective. It should be
noted that the recoil-induced effect (\ref{figuremerit}) is
larger than the  effects discussed in~\cite{ford},
which are related to
metric perturbations associated with the squeezed coherent states
relevant to particle creation in conventional local field theories.
The recoil effects discussed here
are also related to particle creation, but in a string
soliton context~\cite{kanti}.   

We close by recalling that the energy-dependent departures from
an absolute velocity of light that have been discussed here are
amenable to experimental probes. The most promising may involve
observations of distant astrophysical objects whose
emissions exhibit structures within short times $\delta t$~\cite{nature}, 
such as Gamma-Ray Bursters (GRBs), pulsars and active galactic nuclei
(AGNs), with a premium on measurements of the arrival times of
high-energy
photons, so as to maximize the figure of merit $M_{QG} \equiv L \times E /
\delta t$. Past and present observations of such sources have all been
shown to be sensitive to $M_{QG} \sim 10^{16}$~GeV~\cite{test}, within a
few
orders of
magnitude of the Planck mass $M_P \sim 10^{19}$~GeV. We may hope
that future generations of $\gamma$-ray telescopes with larger collection
areas (so as to extend the reach in energy E) and accurate timing (so as
to minimize $\delta t$), such as AMS, GLAST and VERITAS, will be able
to benefit from future identifications of high-redshift sources (so as to
maximize $L$), pushing the experimental sensitivity into the uncharted
territory of quantum gravity.

The work of N.E.M. is supported in part by a P.P.A.R.C. advanced
research fellowship, and that of D.V.N. by D.O.E.
grant DE-FG03-95-ER-40917. We gratefully acknowledge the support
of Hans Hofer and the interest of Marta Felcini.

\end{document}